\documentclass[11pt]{iopart}
\input{amssym}
\usepackage{epsfig}
\global\arraycolsep=1pt

\newcommand{\be}{\begin{equation}}

\newcommand{\ee}{\end{equation}}
\newcommand{\bea}{\begin{eqnarray}}
\newcommand{\eea}{\end{eqnarray}}
\newcommand{\ben}{\begin{equation*}}
\newcommand{\een}{\end{equation*}}
\newcommand{\bean}{\begin{eqnarray*}}
\newcommand{\eean}{\end{eqnarray*}}

\begin{document}
\title{On the accuracy of the PFA: analogies between Casimir 
and electrostatic forces}
\author{Francisco D. Mazzitelli, Fernando C. Lombardo}
\address{Departamento de F\'\i sica {\it Juan Jos\'e
Giambiagi}, Facultad de Ciencias Exactas y Naturales, UBA;
Ciudad Universitaria, Pabell\' on I, 1428 Buenos Aires, Argentina}

\author{Paula I. Villar}
\address{Departamento de F\'\i sica {\it Juan Jos\'e
Giambiagi}, Facultad de Ciencias Exactas y Naturales, UBA;
Ciudad Universitaria, Pabell\' on I, 1428 Buenos Aires, Argentina and 
Computer Applications on Science and Engineering Department, 
Barcelona Supercomputing Center (BSC), 
29, Jordi Girona
08034 Barcelona,
Spain}


\begin{abstract}
We present an overview of the validity of the Proximity Force Approximation (PFA) 
in the calculation of Casimir forces between perfect conductors for different geometries, with particular
emphasis for the configuration of a cylinder in front of a plane. In all cases we compare the exact 
numerical results with those of PFA, and with asymptotic expansions that include  the next to leading order corrections. We also discuss the similarities and differences
between the results for Casimir and electrostatic forces.
\end{abstract}



\newcommand{\beq}{\begin{equation}}
\newcommand{\eeq}{\end{equation}}
\newcommand{\dalam}{\nabla^2-\partial_t^2}
\newcommand{\mbf}{\mathbf}
\newcommand{\itm}{\mathit}
\newcommand{\beqa}{\begin{eqnarray}}
\newcommand{\eeqa}{\end{eqnarray}}

\section{Introduction}
\label{intro}

The experimental and theoretical activity in the analysis of the Casimir effect is, nowadays,
extremely intense. After 60 years, there are several high precision experiments and theoretical calculations for a variety of geometries. In the last years,  there 
has been a
remarkable progress in this field. On the experimental side,
the new generation of experiments started about ten years ago \cite{exp1}. The precision achieved, much
larger than that of the first generation of experiments \cite{exp2}, triggered a lot of theoretical activity.
While there were exact calculations for single cylindrical \cite{self cyl} and spherical 
\cite{self sphere}, perfectly conducting shells,
the calculation of the interaction of two different bodies, beyond the original two parallel plates, 
started about eight years ago. Since then, various theoretical techniques have been developed in order to understand the geometric dependence of the Casimir force. These include the use of the argument theorem to perform explicitly the sum over modes \cite{saharian,Mazzitelli2003,PRARC,NJP},
semiclassical and optical  approximations \cite{semiclass}, methods based on functional integrals
\cite{funct} and scattering theory \cite{scatt}. Many of these approaches have a common root
in the multiple scattering theory developed in the seventies \cite{Balian} (see also \cite{milton wagner} for an updated review and applications to semitransparent bodies), and 
the  evolution in the computational power allowed  a 
 precise numerical evaluation that involves, in general, the computation of determinants of
 infinite matrices.  
There are also full numerical approaches, as the worldline
numerics \cite{gies},  that has been applied to scalar fields satisfying Dirichlet boundary 
conditions, or finite difference methods that evaluate the Casimir energy from the 
two point function of the electromagnetic field \cite{Rodrigues}. 
As a consequence of this theoretical activity, we 
now have exact results for a variety of geometries that involve perfectly conducting shells: 
cylinder and sphere in front of  a plane \cite{Emig,emig,paulo},
eccentric cylinders \cite{PRARC,NJP}, two spheres \cite{emig}, surfaces with periodic corrugations \cite{emig3}, Casimir pistons \cite{pistons}, etc. Some of these methods 
also apply to the case of imperfect mirrors, that we will not consider here.
 
For more than fifty years, the interaction between different bodies was computed mainly using the so called proximity force approximation (PFA) \cite{derjaguin}. This approximation, expected to be valid as long as  the interacting surfaces are smooth and very close, uses the original Casimir expression for  
the energy per unit area for
parallel plates separated by a distance $d$
\beq
E_{\rm pp}(d)=-\frac{\pi^2 }{720 d^3},
\eeq
and approximates the interaction between two conducting surfaces that form a curved  gap of 
variable width $z$ by

\beq
E_{\rm PFA}=\int_\Sigma d\sigma\, E_{\rm pp}(z) .
\label{sigma}\eeq
It is clear that this formula does not take into account the non-parallelism of the 
surfaces. Moreover, the result will depend on the particular surface $\Sigma$ chosen
to perform the integral. However, these corrections are expected to be small for 
low-curvature, very close surfaces.

 Until the development of the theoretical methods described above , the accuracy of the PFA was not assesed, simply because PFA is an uncontrolled approximation, and there were no exact calculations to compare with. On general grounds, denoting by $\mathcal{ L}$ a typical length associated to the curvature of one of the surfaces (assumed much smaller than the curvature of the second surface) and by $d$ the minimum distance between surfaces, one expects 
 \beq
 E_{12}=E_{\rm PFA}\left\{1+\Gamma\frac{\mathcal{L}}{d}+O\left[\left(\frac{\mathcal{L}}{d}\right)^2\right]\right\},  
 \eeq
where $\Gamma$ is a constant, whose numerical value fixes the accuracy of the PFA
in each particular geometry (one can write similar expressions for geometries that involve two surfaces of 
similar curvature).  As we will see, the situation is a bit more complex, since the 
corrections to PFA may contain non-analytic corrections as
$\left(\frac{\mathcal{L}}{d}\right)^n\ln\left(\frac{\mathcal{L}}{d}\right)$.

In this paper, we will present an overview of the accuracy of the PFA for the case of perfectly conducting shells with different geometries: concentric cylinders (Section 2), concentric spheres (Section 3), a cylinder in front of a plane (Section 4), and a 
sphere in front of a plane (Section 5). In all cases, we will compare the exact numerical results with the PFA, and obtain the numerical value of the constant $\Gamma$, which fixes the magnitude of the next to leading 
order (NTLO) correction. Moreover, we will also present, for each geometry, analogous comparisons for the electrostatic energy. These are, of course, trivial textbook
examples. 
However, we think that the computation of the electrostatic energy using PFA is an interesting 
pedagogical exercise that illustrates the accuracy of the approximation for a different interaction, based
on the result for the electrostatic energy contained
between two parallel plates at a potential difference $V$
\beq
U_{\rm pp}=\frac{\epsilon_0 AV^2}{2d}.
\label{upp}
\eeq
Moreover, as we will also point out in our final remarks (Section 6), analogies with classical electromagnetism may be useful to suggest and/or to understand new effects in Casimir physics.

Some of the results presented here have been previously obtained by the authors and collaborators
(concentric cylinders \cite{Mazzitelli2003}, cylinder in front of a plane \cite{NUM}). The exact formula for the Casimir energy in the  concentric-spheres 
geometry has been derived in Ref.\cite{saharian,saharian robin}. However, a  
numerical analysis and a discussion of the relevant limiting situations (in particular the proximity
limit) has not been considered before. Therefore, in Section 4 we describe with some detail the derivation of analytic results in the small and large distance limits,  along with numerical  computations.  For the sake of completeness, we also describe briefly the results
for the sphere-plane configuration obtained by other authors \cite{emig,paulo} (Section 5). 

\section{Concentric Cylinders}
\label{concentric}

Let us first consider two concentric cylinders of length $L$, with radii $a$ and $b$, 
respectively (with $L \gg a, b$ to neglect border effects). The exact formula 
for the Casimir interaction energy is 
given by \cite{saharian,Mazzitelli2003}

\begin{equation}
E_{\rm 12}^{\rm cc} = {L \over 4\pi a^2} \int_{0}^{\infty} d\beta \
\beta\ln M^{\rm cc}(\beta), \label{cc}
\end{equation}
where
\begin{equation}
M^{\rm cc}(\beta)=\prod_n \left[1-{I_n(\beta)K_n(\alpha
\beta)\over I_n(\alpha \beta)K_n(\beta)}\right]
\left[1-{I'_n(\beta)K'_n(\alpha \beta) \over I'_n(\alpha
\beta)K'_n(\beta)}\right] ,\label{Mcc}
\end{equation}
where $\alpha = b/a$. The first factor corresponds to Dirichlet (TM) modes and the
second one to Neumann (TE) modes. The concentric-cylinders
configuration is interesting from a theoretical point of view,
since it can be used to test analytic and numerical methods. It
also has potential implications for the physics of nanotubes
\cite{PRARC,klim}. This result can also be derived as a particular case from the 
general formula for eccentric cylinders \cite{PRARC,NJP}.

The short distance limit $\alpha - 1\ll 1$ has already been analyzed
for this case \cite{Mazzitelli2003}, and involves the summation over all values of $n$,
that can be performed after using the uniform expansion for Bessel functions
(in the next section, we will present a similar calculation for 
concentric spheres).
As expected, the resulting value is equal to the one obtained via
the proximity approximation, namely
\begin{equation}
E_{\rm PFA}^{\rm cc} = - \frac{\pi^3 L}{360 a^2} \;
\frac{1}{(\alpha-1)^3}. \label{proxi}
\end{equation}
When obtaining the PFA for a given configuration, the result is in general ambiguous, 
since it depends on the choice of the surface $\Sigma$ (Eq.(\ref{sigma})). Eq.(\ref{proxi}) corresponds 
to the energy per unit area for parallel plates times the area of the 
{\it inner} cylinder. In this case one could also choose, for instance, the area of the {\it outer} cylinder, which results in
an extra factor of $\alpha$ that modifies the NTLO correction.
The intermediate choice of the {\it geometric mean} of the areas gives
\begin{equation}
E_{{\rm PFA}}^{\rm cc} = - \frac{\pi^3 L}{360 a^2} \;
\frac{\alpha^{1/2}}{(\alpha-1)^3}, \label{proxigeo}
\end{equation}
and reproduces the result that is obtained using a semiclassical approximation based on
periodic orbit theory  \cite{Mazzitelli2003}.

In the opposite limit ($\alpha \gg 1$), it can be shown
that to leading order only the TM $n=0$ mode contributes to
the interaction energy, and that the energy decreases logarithmically
with the ratio $\alpha = b/a$,
\begin{equation}
E_{12}^{\rm cc} \approx - {1.26 L \over 8\pi b^2\ln\alpha}.
\label{cclargea}
\end{equation}
It is worth to stress that, while for small values of $\alpha$ both
TM and TE modes contribute with the same weight to the interaction
energy, the TM modes dominate in the large $\alpha$ limit.

In previous works, we have evaluated the analytic corrections to the PFA
given in Eq.(\ref{proxi}). Due to the simplicity of this configuration, 
it is possible to obtain not only the next to
leading order, but also the next to next to leading contribution \cite{NUM,JPA}.
The Casimir energy, beyond the
proximity approximation, can be written as \cite{NUM,JPA}
\begin{equation}
E_{12}^{\rm cc} \approx -\frac{\pi^3 L}{360 a^2(\alpha -
1)^3}\left\{1 + \frac{1}{2}(\alpha - 1) - (\frac{1}{10}+\frac{2}{\pi^2})(\alpha -1)^2+...\right\}
.\label{ntntlead}\end{equation}
In the expression above, the first term inside the parenthesis
corresponds  to the proximity approximation contribution in
Eq.(\ref{proxi}), while the second and third terms are the first
and second order corrections respectively. It is important to
stress here that both TM and TE modes contribute with the same
weight to the energy up to the next to leading order, but it is
not the case in the second order correction \cite{NUM,JPA}. 
It is also remarkable that the PFA based on the geometric mean of 
the areas given in Eq.(\ref{proxigeo}) reproduces the exact result not only to leading order 
but also to the  NTLO. In Refs.\cite{NUM,JPA} we have shown that PFA can be used as a useful tool in 
order to improve the numerical evaluation at very small distances, and we have used this improvement in order
to check numerically the non linear correction to PFA described in Eq.(\ref{ntntlead}).

Let us now consider the electrostatic 
analogue for this configuration. It is trivial to evaluate the 
exact expression for the electrostatic interaction energy, which can be
written as
\begin{equation}
 U_{12}^{\rm cc} = \frac{\pi \epsilon_0 V^2}{\ln\alpha},
 \label{uexactcc}
\end{equation}
where $V$ is the difference between the electrostatic potential of 
the inner and outer cylinders. 

The proximity approximation 
for the electrostatic interaction energy can be computed from 
the result of two parallel plates Eq.(\ref{upp}), and it is
given by
\begin{equation}
 U_{\rm PFA}^{\rm cc} \approx \frac{\pi \epsilon_0 V^2}{\alpha - 1},
\end{equation}
where we have used the area of the inner cylinder.
Taking the ratio between the exact and PFA results , it is possible to read the 
next to leading correction, which is given by
\begin{equation}
 \frac{U_{12}^{\rm cc}}{U_{\rm PFA}^{\rm cc}}\approx 1 + \frac{1}{2} \left(\alpha - 1\right).
\end{equation}

Remarkably, the NTLO correction has the same numerical factor both
for the electrostatic interaction energy  and for the Casimir interaction energy
shown in Eq.(\ref{ntntlead}). Related to this,  the calculation of $ U_{\rm PFA}^{\rm cc}$ 
using the geometric mean of the areas also reproduces the exact result
 $U_{12}^{\rm cc}$ including the NTLO.
 
 There is an additional analogy between the calculations of the electrostatic and Casimir
 energies: in the large distance limit $\alpha\gg 1$, both energies vanish only logarithmically
 as the radius
 of the inner cylinder tends to zero (see Eqs.(\ref{cclargea}) and (\ref{uexactcc})). 
 
\section{Concentric spheres}
\label{cs}

Let us now consider two concentric spherical shells of radii $a$ and $b$ respectively, with $\alpha =b/a>1$. 
The Casimir interaction energy can be computed using a procedure similar to that of the concentric cylinders. 
The exact energy is given 
by \cite{saharian,saharian robin}
\beq
E_{12}^{\rm cs} = \frac{1}{\pi a}\sum_{l\geq 1}\nu\int_0^\infty dy\,  \ln[(1-F_\nu^{\rm TE})(1-F_\nu^{\rm TM})]\,\, ,
\label{espheres}
\eeq
where 
\beq
F_\nu^{\rm TE}=\frac{I_\nu(y)K_\nu(\alpha y)}{I_\nu(\alpha y)K_\nu(y)}\,\, ,
\label{FTE}\eeq
\beq
F_\nu^{\rm TM}=\frac{(I_\nu(y)+2yI'_\nu(y))(K_\nu(\alpha y)+2\alpha y K'_\nu(\alpha y))}{(I_\nu(\alpha y)+2\alpha y I'_\nu(\alpha y))(K_\nu(y)+2 y K'_\nu(y))}\,\, ,
\label{FTM}\eeq
and $\nu=l+1/2$.

As far as we know, this energy has not been studied in detail before, so we 
analyze the opposite limits $\alpha\rightarrow 1$ and $\alpha\rightarrow\infty$ . 
In order to obtain an analytic expression in
the proximity limit
$\alpha\rightarrow 1$, it is useful to perform the change of variables $y\rightarrow \nu y$ in the integral 
appearing in Eq.(\ref{espheres}), so we can use the uniform expansion for the Bessel functions. For example 
we have
\begin{equation}
\frac{K_\nu(\nu \alpha y)}{K_\nu(\nu y)} = \frac{\sqrt{1 + y^2}}{\sqrt{1 +
\alpha^2 y^2}}\frac{(1 - \frac{u(t_\alpha )}{\nu})}{(1 -
\frac{u(t_1)}{\nu})} e^{\nu\left[\eta (\alpha y) - \eta (y)\right]},
\end{equation}
where
\begin{equation}
\eta (y)= \sqrt{1 + y^2}+ \ln{\frac{y}{1 + \sqrt{1 + y^2}}} ~;~
u(t) = \frac{3 t - 5 t^3}{24}~;~ t_\alpha = \frac{1}{\sqrt{1 +
\alpha^2 y^2}},
\end{equation}
and similar expressions for the functions $I_\nu, I'_\nu,$ and $K'_\nu$. Inserting these asymptotic expansions
in Eqs.(\ref{FTE}) and (\ref{FTM}), one can show that
\beq
F_\nu^{\rm TE}\simeq F_\nu^{\rm TM}\simeq e^{-2\nu\Delta\eta(y)}(1+O(\frac{\alpha -1}{\nu})),
\eeq
where 
\beq
\Delta\eta(y)=\eta(\alpha y)-\eta(y)\simeq (\alpha -1)\sqrt{1+y^2}-\frac{(\alpha - 1)^2}{2\sqrt{1+y^2}}\,\, .
\eeq
The term proportional to $(\alpha -1)/\nu$ will not contribute to the leading and NTLO,
so we will neglect it in what follows.

Using these expressions, we can write the interaction energy as
\beq
E_{12}^{\rm cs}\simeq -\frac{2}{\pi a}\int_0^\infty dy\,\sum_{k\geq 1}\frac{1}{k}\sum_{l\geq 1}\nu^2 e^{-2\nu k \Delta\eta(y)}
.\label{espheresapprox}
\eeq
The sum over $l$ can be easily computed and gives
\bea
&\sum_{l\geq 1}&\nu^2 e^{-2\nu k \Delta\eta(y)} = \frac{1}{4 k^3\Delta\eta^3}+O(\Delta\eta)\nonumber\\
&=&   \frac{1}{4 k^3(\alpha - 1)^3(1+y^2)^{3/2}}\left(1+\frac{3(\alpha - 1)}{1+y^2}+O((\alpha -1)^2)\right)
. \label{sum}
\eea
Inserting Eq.(\ref{sum}) into Eq.(\ref{espheresapprox}), computing first the sum over $k$
and then the remaining integral we finally obtain
\beq
E_{12}^{\rm cs}=E_{\rm PFA}^{\rm cs}\left\{1 + (\alpha - 1) + O((\alpha - 1)^2)\right\}. 
\label{espheres final}
\eeq
Here
\beq
E_{\rm PFA}^{\rm cs} = -\frac{\pi^2}{720 (b-a)^3} 4\pi a^2 = -\frac{\pi^3}{180 a (\alpha -1)^3}, 
\eeq
is the Casimir energy computed with the PFA using the area of the inner surface. 
We have confirmed the analytic approximation given in Eq.(\ref{espheres final}) through a numerical evaluation of the exact
energy given Eq.(\ref{espheres}). The results are shown in Fig. 1.

There are some interesting properties, similar to those of the previous section,  that are worth noticing. On the 
one hand, TE and TM modes give the same contribution to both the leading and next to leading orders. On the other
hand, if the PFA approximation is computed with the geometric mean area of the inner and outer spheres, the resulting 
expression
\beq
E_{\rm PFA}^{\rm cs}= -\frac{\pi^3\alpha}{180 a (\alpha -1)^3}=-\frac{\pi^3}{180 a (\alpha -1)^3}\left\{1 + (\alpha - 1)\right\},
\eeq
reproduces not only the leading term of the exact interaction energy but also 
the NTLO.

\begin{figure}[!ht]
\centering
\includegraphics[width=14cm]{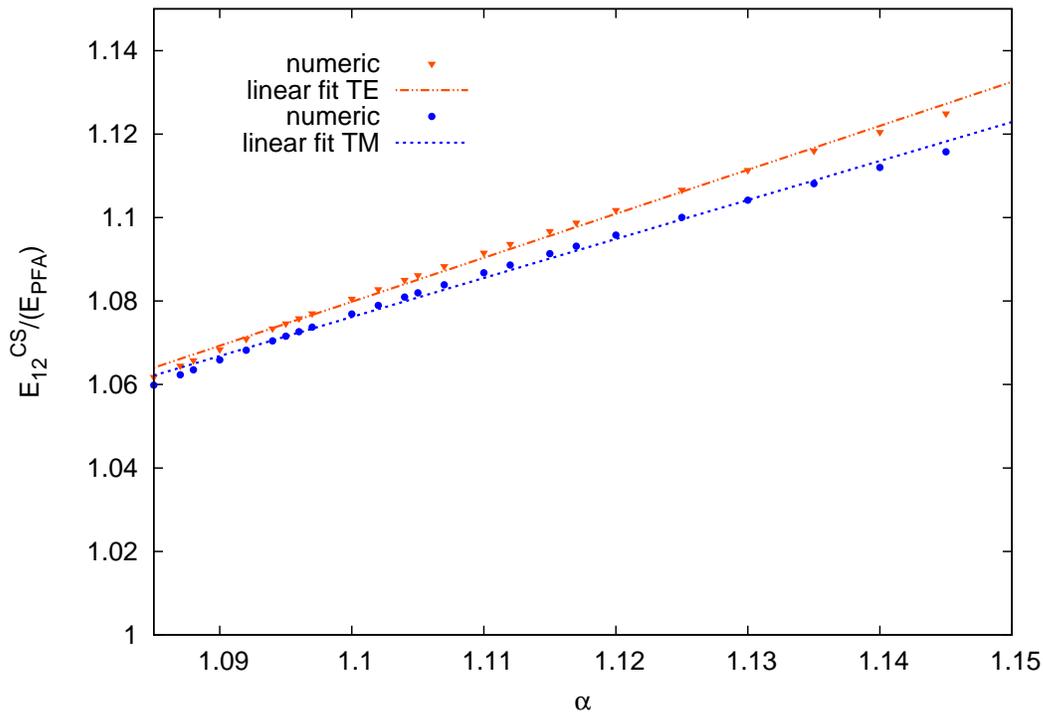}
\caption{Numerical evaluation of the Casimir interaction energy for the
configuration of concentric spheres,
near the proximity limit. A simple fit $f(x) = a + b x$ of the numerical data gives, 
for the TM-modes $a=0.97,\, b=1.09 $, and $a=0.98,\, b=0.97$ for TE-modes } \label{csphere}
\end{figure}

\begin{figure}[!ht]
\centering
\includegraphics[width=14cm]{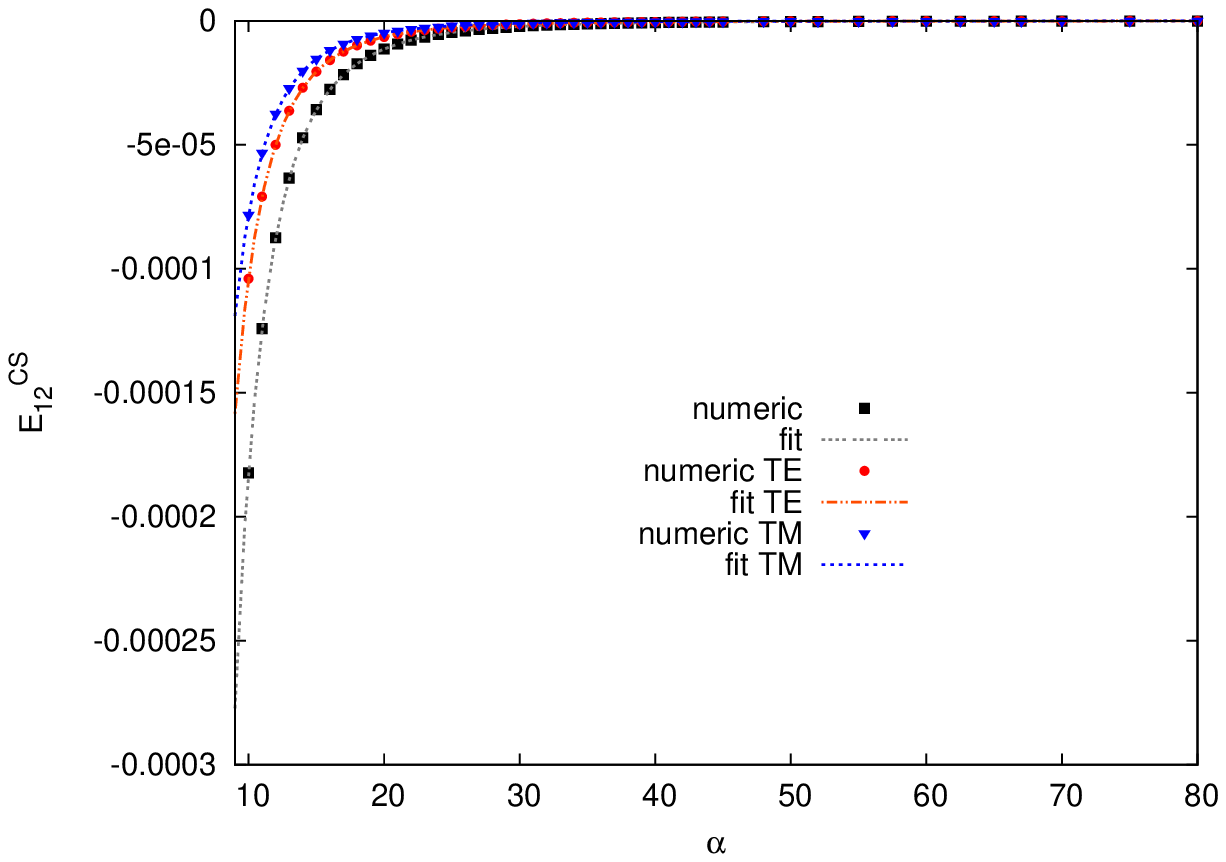}
\caption{Large $\alpha$ behaviour of the Casimir interaction energy for concentric 
spheres. A simple fit of the form $f(x)= a/x^b$ gives $a=-1.05,\, b=-4.01$ for TE modes,
and  $a=-0.82,\, b=-4.02$ for TM modes.} \label{csphere3}
\end{figure}

One can also study the opposite limit, in which $a\ll b$. In this case, the Casimir interaction energy is dominated
by the $l=1$ modes. Keeping only this contribution in the exact expression Eq.(\ref{espheres}),
and after the change of variables $\alpha y=x$ we obtain
\beq
E_{12}^{\rm cs,TE}\approx\frac{3}{2\pi a\alpha}\int_0^\infty dx\, \ln\left (1-\frac{K_{3/2}(x)I_{3/2}(x/\alpha)}{I_{3/2}(x)K_{3/2}(x\alpha)}\right )
.\label{l=1}
\eeq
Expanding the logarithm up to the leading order
in $1/\alpha$ we get
\beq
E_{12}^{\rm cs,TE}\approx-\frac{1}{\pi^2 a\alpha^4}\int_0^\infty dx\, x^3\frac{K_{3/2}(x)}{I_{3/2}(x)}
\approx -\frac{0.745}{a \alpha^4}.
\label{alphagrandeTE}
\eeq
A similar analysis can be carried out for the TM modes. The result is 
\beq
E_{12}^{\rm cs,TM}\approx -\frac{1.011}{a \alpha^4}.
\label{alphagrandeTM}
\eeq
It is interesting to note that, unlike the case of concentric cylinders, both TE and TM modes contribute with the same order of magnitude to
the Casimir energy in the large $\alpha$ limit. Moreover the interaction energy vanishes as
$a^3$ as the radius of the sphere tends to zero. We have checked these analytic
results with numerical evaluations of the exact formula, as shown in Fig.2. A fit of the form 
$f(x)= a/x^b$ gives $a=-1.05,\, b=-4.01$ for TM modes, and  $a=-0.82,\, b=-4.02$ for TE modes. Moreover, 
performing a fit with $g(x) = a/x^4$, we obtain $a= -1.03$ and $a= -0.78$ for TM and TE modes respectively.

Let us now compare the PFA in Casimir physics with the textbook
electrostatic example. The electrostatic energy contained between the spherical shells
is given by
\beq
U_{12}^{\rm cs}=2\pi\epsilon_0 V^2 \frac {ba}{b-a},
\eeq
where $V$ is the potential difference.
A trivial application of the PFA, based on the inner sphere, gives
\beq
U_{\rm PFA}^{\rm cs}=2\pi\epsilon_0 V^2 \frac {a^2}{b-a},
\eeq
so 
\beq
U_{12}^{\rm cs}=U_{\rm PFA}^{\rm cs}\frac {b}{a}=U_{\rm PFA}^{\rm cs} \left\{1+(\alpha -1)\right\}. 
\eeq
We see that, as for the case of concentric cylinders, the next to leading order correction to 
PFA has the same numerical coefficient in electrostatic and Casimir energies. Moreover, in this case,
the choice of the geometric mean area gives the exact result for the electrostatic energy.
As we will see 
in the next sections, these are peculiarities of the geometries considered so far.

\section{A cylinder in front of a plane}

We consider now a perfectly conducting cylinder of length $L$ and radius $a$ 
(with $L\gg a$ to neglect border effects). The cylinder is parallel  to a perfectly conducting 
planar surface of area $A\gg a^2$, and the minimum distance between the two surfaces is 
denoted by $d$. This configuration is  of experimental
interest: being intermediate between the sphere-plane and the
plane-plane geometries,  it can shed some light on the
longstanding controversy about thermal corrections to the Casimir
force. Keeping the two plates parallel has proved very difficult,
while the sphere and plate configuration avoids this problem, the
force is not extensive. In the case of the cylinder-plane
configuration, it is easier to hold the cylinder parallel and the
force results extensive in its length. There is an ongoing
experiment to measure the Casimir force for this configuration \cite{hayespra72}.

The Casimir energy for this configuration was first
evaluated in the PFA in Ref.\cite{europhysics}. The exact formula
has been derived in Refs.\cite{Emig, Bordag}, and has the same
structure than Eq.(\ref{cc}), where

\begin{eqnarray}
 E_{12}^{\rm cp} &=&  \frac{L}{4 \pi  a^2} \int_0^{\infty}  d\beta ~\beta
~\left[{\rm ln}(M^{\rm TE}(\beta))
+ {\rm ln}(M^{\rm TM}(\beta))\right]  \nonumber \\
&=& E^{\rm TE}_{12}+E^{\rm TM}_{12} , \label{cp}
\end{eqnarray}
where $M^{\rm TM}(\beta)={\rm det}[\delta_{n p}- A_{n,p}^{\rm
TM, CP}]$ and $M^{\rm TE}(\beta)={\rm det}[\delta_{n p}- A_{n,p}^{\rm
TE, CP}]$. Here $\beta $ is a dimensionless integration variable and
$n,p$ are arbitrary integers. The matrix elements are given by \cite{Emig,Bordag}

\begin{eqnarray}
A_{n,p}^{\rm TE, CP} &= & - \frac{I'_n(\beta)}{K'_n(\beta)}
K_{n+p}(2 \beta H/a),
\label{AcpTE}
\end{eqnarray} and
\begin{eqnarray}
A_{n,p}^{\rm TM, CP} &= & \frac{I_n(\beta)}{K_n(\beta)}
K_{n+p}(2 \beta H/a).
\label{AcpTM}
\end{eqnarray}
Note that the evaluation of the Casimir energy for this configuration 
involves the computation of the determinant of an infinite, non-diagonal
matrix. Once more, the exact formula can be derived from the general formula 
for eccentric cylinders \cite{NJP}.

\begin{figure}[!h]
\centering
\includegraphics[width=14cm]{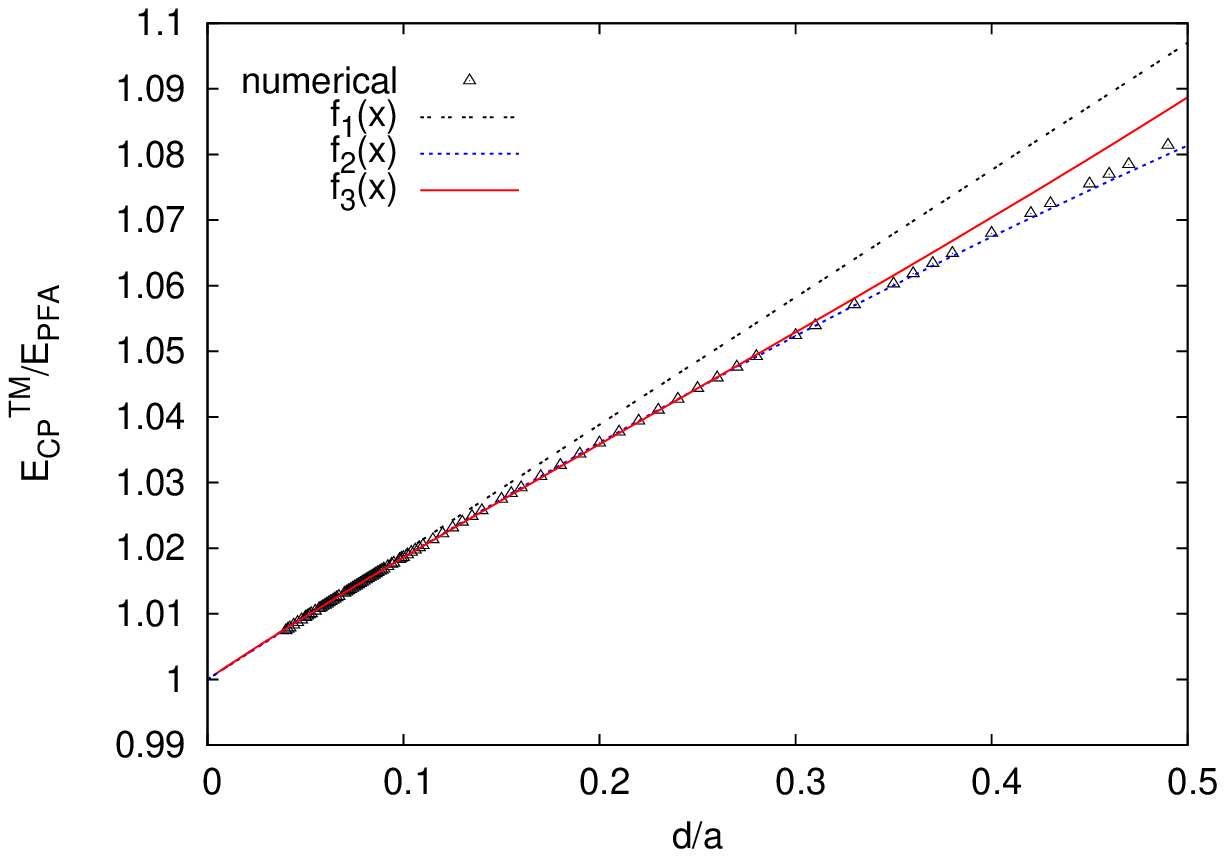}
\caption{Numerical result for the TM modes
for the cylinder-plane configuration, and the
corresponding fits presented in Table \ref{tabla1}. A simple
linear fit $f(x)= a+ b x$ of the numerical data in the
interval $0.04 \leq d/a \leq 0.07$ gives $a = 0.9999$ and
$b = 0.1900$. The theoretical values are $a = 1$ and $b = 0.1944$.}
\label{CPTM}
\end{figure}

In the following we will numerically evaluate the cylinder-plane Casimir interaction 
energy for small distances, in order to discuss the leading correction to the PFA.
In Figs.\ref{CPTM}, \ref{CPTE}, and \ref{CPTM2}
we present the Casimir interaction energy for the cylinder-plane
configuration. For the runs, we used a matrix of dimension (101,101) to reach the
proximity limit ($d\rightarrow 0$). It must be mentioned that for
smaller values of $d$, we need to increase the dimension of the A
matrix and the integration range of $\beta$ in Eq.(\ref{cp}).
This fact becomes our major limitation to reach yet smaller values
of $d$.

This problem has
been considered from an analytical point of view in
Ref.\cite{Bordag}. Using the uniform expansions for the Bessel
functions appearing in the matrix elements $A_{n,p}^{\rm TE, CP}$
and $A_{n,p}^{\rm TE, CP}$, and after complex calculations, it can
be shown that, in the proximity limit:
\begin{equation}
E_{12}^{\rm cp,TM}=-\frac{1}{2\pi}\sqrt{\frac{a}{d^5}}\frac{3\zeta(4)}{32\sqrt{2}}
\bigg( 1+ 0.1944 {\frac{d}{a}} + ... \bigg),
\label{E12CPBTM}
\end{equation}
\begin{equation}
E_{12}^{\rm cp,TE}=-\frac{1}{2\pi}\sqrt{\frac{a}{d^5}}\frac{3\zeta(4)}{32\sqrt{2}}
\bigg( 1 - 1.1565 {\frac{d}{a}} + ... \bigg),
\label{E12CPBTE}
\end{equation}
where we have written separately the contributions of TM and TE modes.

\begin{center}
\begin{table}
\begin{tabular}{c|c|c|c}
  $d/a$   &  $f_1(x)=1+b x$  & $f_2(x)=1+b*x+c*x^2$   & $f_3(x)= 1+b*x+c*x^2*\log(x)$   \\  \hline \hline
$[0.04:0.15]$ & $b = 0.1864$ & $b = 0.1922, c= -0.0601$ & $b= 0.1961, c=0.0438$ \\ \hline
$[0.04:0.20]$ & $b= 0.1849$ & $b= 0.1923, c= -0.0613$ & $b= 0.1983, c=0.0540$ \\  \hline
$[0.04:0.25]$ & $b= 0.1829$ & $b= 0.1922, c= -0.0601$ & $b= 0.2003, c=0.0634$ \\ \hline
$[0.04:0.30]$ & $b= 0.1811$ & $b= 0.1920, c= -0.0586$ & $b= 0.2022, c= 0.0716$ \\ \hline
$[0.04:0.35]$ & $b= 0.1794$ & $b= 0.1918, c= -0.0572$ & $b= 0.2045, c= 0.0810$ \\ \hline
$[0.04:0.40]$ & $b= 0.1771$ & $b= 0.1914, c= -0.0549$ & $b= 0.2076, c= 0.0935$ \\ \hline
 \hline
\end{tabular}\caption{Different fits for the numerical results of Fig. \ref{CPTM} (TM modes).
We fix $f_i (0) = 1$ since the numerical data agree this value
with high precision.}\label{tabla1}
\end{table}

\begin{figure}[!ht]
\centering
\includegraphics[width=14cm]{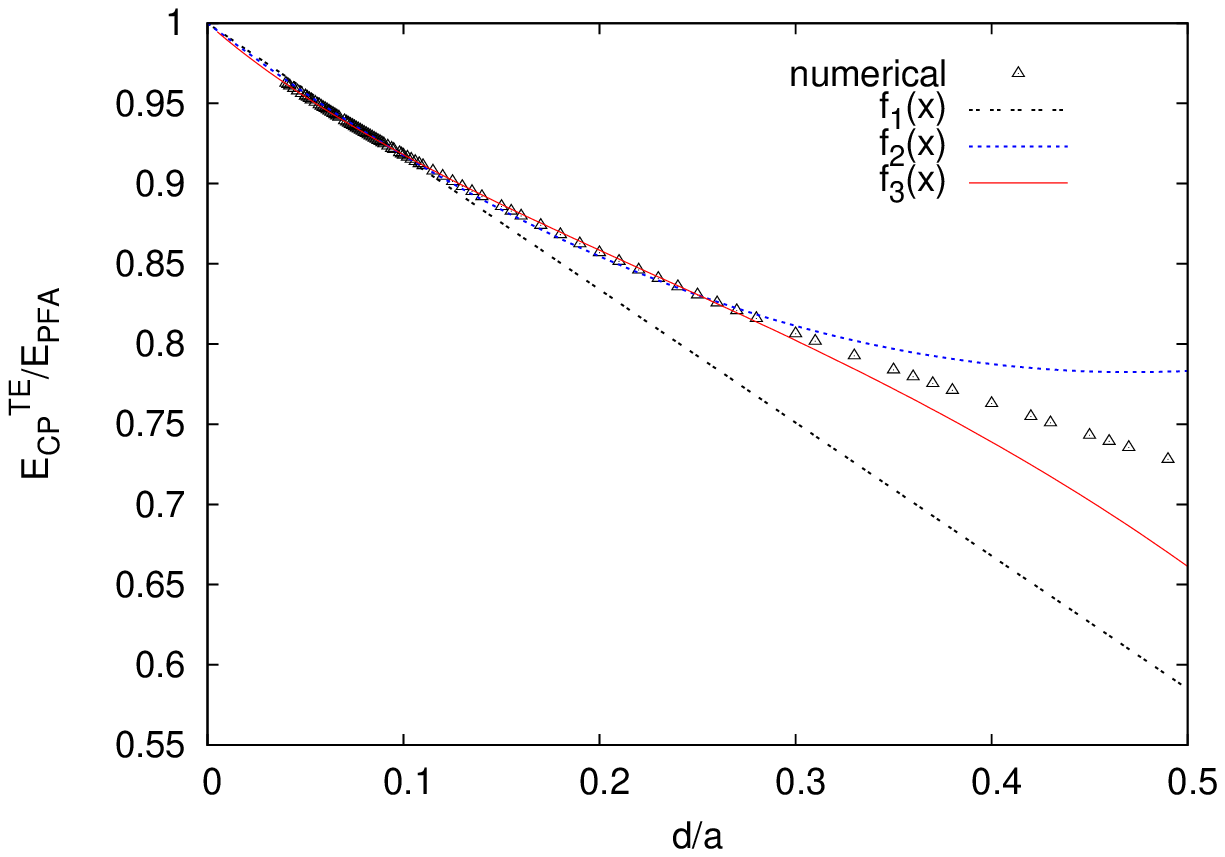}
\caption{ Numerical result for the TE modes
for the cylinder-plane configuration, and the
corresponding fits presented in Table \ref{tabla2}. A simple
linear fit $f(x)= a+ b x$ of the numerical data in the
interval $0.04 \leq d/a \leq 0.07$ gives $a = 0.9940$ and
$b = - 0.7808$. The theoretical values are $a = 1$ and $b = -1.1565$.}
\label{CPTE}
\end{figure}

\begin{figure}[!ht]
\centering
\includegraphics[width=14cm]{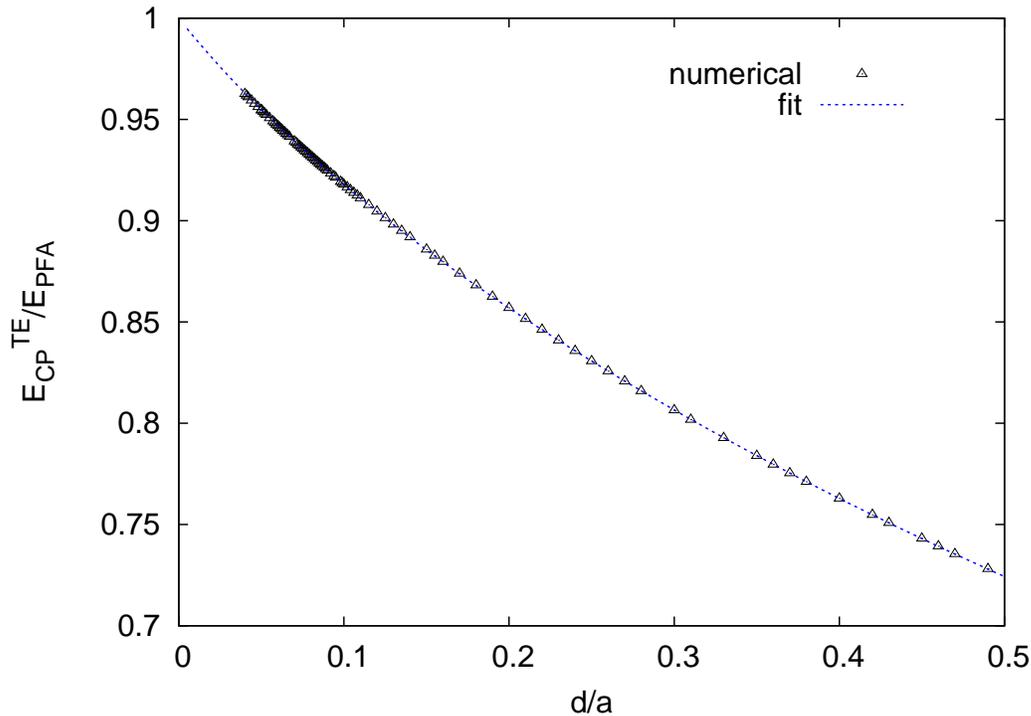}
\caption{A numerical fit of the results for the TE modes including
cubic corrections $f(x)= 1 + b x + c x^2 \log x + d x^3$. The
coefficients are $b=-1.0478$, $c= - 0.9485$, and $d = 0.6708$.}
\label{CPTM2}
\end{figure}

\begin{table}
\begin{tabular}{c|c|c|c}
  $d/a$   &  $f_1(x)=1+b x$  & $f_2(x)=1+b*x+c*x^2$   & $f_3(x)=1+b*x+c*x^2*\log(x)$   \\  \hline \hline
$[0.04:0.15]$ & $b = -0.8301$ & $b = -0.9704, c =  1.4499$ & $b = -1.0711 , c = -1.0852$ \\ \hline
$[0.04:0.20]$ & $b = -0.8013$ & $b = -0.9509 , c =  1.2326$ & $b = -1.0772 , c = -1.1141$ \\ \hline
$[0.04:0.25]$ & $b = -0.7683$ & $b = -0.9349 , c = 1.0794 $ & $b = -1.0890 , c = -1.1674$ \\ \hline
$[0.04:0.30]$ & $b = -0.7399$ & $b = -0.9222, c = 0.9772 $ & $b = -1.1037, c = -1.2306$ \\ \hline
$[0.04:0.35]$ & $b = -0.7158$ & $b = -0.9091, c = 0.8879$ & $b = -1.1232, c =  -1.3115$ \\ \hline
$[0.04:0.40]$ & $b = -0.6851$ & $b = -0.8943, c = 0.7999$ & $b = -1.1534, c = -1.4360$ \\\hline
 \hline
\end{tabular}\caption{Different fits for the numerical results of Fig. \ref{CPTE} (TE modes).
We fix $f_i (0) = 1$ since the numerical data agree this value
with high precision.}\label{tabla2}
\end{table}\end{center}

We will discuss the first order corrections to PFA for TM and TE
modes separately. In Fig.\ref{CPTM}, we show our numerical results
for the TM modes.  The fit of the numerical results depends of
course on the interval chosen for $d/a$. There is an obvious
compromise: on the one hand,  as already mentioned, we cannot
consider very small values for  $d/a$ because of numerical
limitations. On the other hand, the expansion in powers of $d/a$
are expected to be valid only for $d/a\ll 1$.   In any case, as
can be seen from Table 1,  the different fits for the numerical
results are stable, and confirm both the PFA to leading and next
to leading orders. Indeed, the results are fully compatible with
the analytic results given in Eq.(\ref{E12CPBTM}), considering
both linear and quadratic fits of the numerical results. Moreover,
a simple linear fit in a smaller range of $d/a$ gives $a = 0.9999$
and $b= 0.1900$ and already reproduces the analytical results
\cite{Bordag} with high accuracy (see also numerical findings in
\cite{gies2}).

In Fig.\ref{CPTE}, we show our results for the Neumann modes, and
we include in Table 2 different fits of the numerical data. In
this case, the value obtained for the linear correction to PFA
depends strongly on the assumption about the next non trivial
correction. This is not surprising: as we cannot consider
extremely small values for $d/a$, the non linear corrections may
have a non  negligible contribution in the intervals chosen for
the fits. For example, a simple linear fit gives $a = 0.994$ and
$b= -0.7808$ which does not coincide with the result in
Eq.(\ref{E12CPBTE}). However, based on the discussion about the
slower convergence of the Neumann corrections presented in
Ref.\cite{Bordag}, we have allowed the possibility of non linear
corrections proportional to $(d/a)^2 \ln(d/a)$ in our fits.
Remarkably, when this non linear corrections are taken into
account, the coefficient of the  linear correction gets closer to
the analytic prediction in Eq.(\ref{E12CPBTE}), that we reproduce
with an error less than $7\%$. Note that, as can be seen in
Fig.\ref{CPTM}, this is not the case for TM modes, since the best
fit of the numerical data contains a quadratic term without a
logarithm. In Fig.\ref{CPTM2} we show a fit of the numerical data for TM
modes that includes a cubic correction $(d/a)^3$. With this
additional term, the fit reproduces the numerical data up to
$d/a=0.5$.

To summarize our results, the fits of the numerical data clearly confirm the
analytic prediction for the TM modes, and suggest that the next
non trivial correction for the TE modes is not quadratic but
proportional to $(d/a)^2 \ln(d/a)$.

As in the previous sections, we consider the electrostatic interaction. 
Let us assume that the conducting cylinder is kept at a fixed electrostatic potential $V$,
while the planar surface is grounded. For this geometry, the exact electrostatic interaction 
energy given by 

\begin{equation}
 U_{12}^{\rm cp} =  \frac{\pi L \epsilon_0 V^2}{{\rm arccosh}\left(1 + \frac{d}{a}\right)}.
\end{equation}

In the limit of $d/a\ll 1$ it is simple to show that the electrostatic energy reduces 
to the PFA result

\begin{equation}
 U_{\rm PFA}^{\rm cp} = \frac{L \pi \epsilon_0 V^2}{\sqrt{2}}\sqrt{\frac{a}{d}}.
\end{equation}

As it was done before, we can compare the exact electrostatic energy (expanded in powers of 
$d/a$) with the PFA result, and extract from it the next to leading correction, i.e., 

\begin{equation}
 \frac{U_{12}^{\rm cp}}{U_{\rm PFA}^{\rm cp}} = 1 + \frac{1}{12} \frac{d}{a}.
\end{equation}

As in the previous examples, the exact electrostatic 
result shows a linear NTLO correction to PFA.  However, the numerical value of 
the linear correction is different from that of Casimir energy.

Finally, we also point out that, in the large distance limit $d\gg a$, the electrostatic interaction 
becomes
\begin{equation}
 U_{12}^{\rm cp} =  \frac{\pi L \epsilon_0 V^2}{\ln(\frac{d}{a})}.
\end{equation}
As for the Casimir energy \cite{Emig}, it vanishes logarithmically with as $a\rightarrow 0$.

\section{A sphere in front of a plane}
\label{sp}
The sphere-plane geometry is, up to now, the most important geometry that have been used to measure precisely the Casimir forces. 
From the theoretical point of view, the evaluation of the Casimir energy in the electromagnetic case has been performed 
very recently in Refs.
\cite{emig,paulo}, while the evaluation for scalar fields has been previously reported in Ref.\cite{scatt}. See
also \cite{bordag sphere} for asymptotic expansions in the scalar field case near the proximity limit. 

For the sake of completeness,
we quote here the results obtained in Refs.\cite{emig,paulo} regarding the behaviour of the Casimir energy
for this configuration. Denoting by $a$ the radius of the sphere, and by $d$ the minimum distance 
between the plane and the sphere, numerical fits in both references give
\beq
E_{12}^{\rm sp}\simeq E_{\rm PFA}^{\rm sp}\left\{1 - 1.4\frac{d}{a}\right\}.
\label{emigpfa}
\eeq
The next to leading order correction is again  linear, as in the previous cases. Both fits
were performed by assuming that the next to NTLO is quadratic in $d/a$.

There is still no
analytic prediction for the NTLO correction in the electromagnetic case. However,  one can compare the
results of the numerical calculations \cite{emig} and the asymptotic expansions in the scalar case \cite{bordag sphere}.
Although the scalar results for TE and TM modes do not reproduce the electromagnetic result (this geometry does not allow
this decomposition), there is an interesting similarity with the results described in the previous section. The theoretical
asymptotic expansions, for scalar fields satisfying Dirichlet and Neumann boundary conditions read, respectively,
\bea 
E_{12}^{\rm sp,D}&\simeq& E_{\rm PFA}^{\rm sp}\left\{1 + \frac{1}{3}\frac{d}{a}\right\},\nonumber\\
E_{12}^{\rm sp,N}&\simeq& E_{\rm PFA}^{\rm sp}\left\{1 + (\frac{1}{3}-\frac{10}{\pi^2})\frac{d}{a}\right\}.
\label{bordagnus}
\eea
The numerical fits for the scalar case \cite{emig} give $0.33$ and $-2.43$, for the Dirichlet and Neumann case, respectively.

While the agreement for Dirichlet modes is remarkable, for Neumann modes there is a strong discrepancy.
So, based on the discussion for the cylinder-plane geometry, one can argue that also in this case the second order corrections could contain logarithmic factors.

The electrostatic problem can also be solved exactly, and it is relevant for the initial calibration in the 
measurements of the Casimir force. If the potential difference between the plane and the sphere is $V$,
the electrostatic energy is given by \cite{smythe}
\beq
U_{12}^{\rm sp}=2\pi\epsilon_0 a V^2 \sinh\beta\sum_{n\geq 1}\frac{1}{\sinh(n\beta)},
\label{uexactsp}
\eeq
where $\cosh\beta = 1+d/a$. In order to obtain an analytic expression in the limit $\beta\rightarrow 0$
we write
\bea
S\equiv \sum_{n\geq 1}\frac{1}{\sinh(n\beta)}&=& S-\int_1^{\infty}\frac{dn}{\sinh(n\beta)}+\frac{1}{\beta}\ln(\coth\beta)\nonumber\\
&=& \frac{\gamma}{\beta} + \frac{1}{\beta}\ln(\coth\beta) + O(\beta),
\label{evalsum}
\eea
where $\gamma = 0.5772$. Replacing this expression into Eq.(\ref{uexactsp}) and expanding the result for small $d/a$
we obtain
\beq
U_{12}^{\rm sp}=U_{\rm PFA}^{\rm sp}\left\{1+\frac{1}{3}\frac{d}{a}+O\left (\frac{d/a}{\ln(d/a)}\right )\right\},
\label{usppfa}\eeq
where 
\beq
U_{\rm PFA}^{\rm sp}=-\pi\epsilon_0 a V^2 \ln (2 d/a)\,\, .
\eeq
In Eq.(\ref{usppfa}) we omited an irrelevant constant term.
It is interesting to remark that the next to NTLO correction in the 
electrostatic force is not quadratic  but proportional to $d/(a \log(d/a))$.

\section{Final remarks}
We have presented a brief review of the calculations of the Casimir energy for different geometries
involving perfect conductors, paying particular attention to the NTLO corrections to the PFA. In all cases considered, the first corrections to the PFA 
are linear, with a coefficient of order one. So, generically,  the PFA results agree with the exact 
energies within $1\%$ when $\mathcal{L}/d < 10^{-2}$. The situation for the next to NTLO is 
more complex. 
For concentric cylinders
this correction is quadratic \cite{NUM,JPA}, and it can be shown that this is also the case for concentric spheres. 
However, in the cylinder-plane configuration additional logarithmic factors could arise \cite{NUM,Bordag}. This is  
probably also 
the case for a sphere in 
front of a plane.

We have compared validity of the PFA for the Casimir interaction energy with the same approximation in electrostatic examples, in each geometry considered.
In all cases, the general result is also valid: the NTLO corrections to PFA are always linear. 
Moreover,
for concentric spheres and cylinders, the next to leading order corrections to PFA have the same
numerical coefficients for electrostatic and Casimir energies, and can be obtained from the PFA using a 
particular area, i.e. the geometric mean of the areas of both surfaces. This is certainly a property of this 
particular geometries, in which the distance between surfaces is constant and the normal to both surfaces are
parallel at each point. 

There is another property of the Casimir interaction energy that has its counterpart in electrostatics. In the case of concentric cylinders or cylinder-plane geometries, the Casimir energy vanishes only logarithmically as the radius $a\rightarrow 0$. 
This is also the case for the analogous electrostatic problems. Once more, this is a property of geometries involving cylinders,
and the situation is different for geometries involving spheres, as we have shown in Section 3.

The analogies between the Casimir energy and the electrostatic energy 
could be useful to suggest and/or understand the behaviour of the vacuum forces
in different situations. Let us consider, instead of
perfect conductors, the case of surfaces
that separate media with different electromagnetic properties. For
example, consider three media described by different dielectric constants $\epsilon_1>\epsilon_2>\epsilon_3$,
separated by flat surfaces. It is a simple exercise to show that, even if the interfaces 
have free electric charges of
different sign,  the interaction between them may be repulsive, due to the polarization of the media.
This suggests
that the same situation may happen for the vacuum fluctuations, and this is indeed the case, as can be easily shown
using Lifshitz formula \cite{lifshitz}. Similar electrostatic effects arise for all the geometries considered here.
Therefore, based on this analogy, one can argue that repulsive Casimir forces can take place in all of these geometries,
if the boundaries become interfaces between different media. This should be valid even beyond
the obvious situation in which one uses the PFA starting from Lifshitz formula. There is a concrete example that has been 
recently analyzed, the repulsive interaction between eccentric cylinders \cite{dalvit08}. According to the electrostatic 
analogy this property should be valid as long as the radii of the cylinders and the dielectric constants satisfy certain 
relations. It would be interesting to check if this is also the case for the Casimir interaction. Work on this issue is in progress. 

\ack
We would like to thank V. Dodonov for his hospitality and organization of the meeting {\it{``60 years of Casimir effect''}}. 
We also thank Mariano Vazquez for 
useful computational help. This work was supported by
UBA, Conicet, and ANPCyT Argentina.

\end{document}